\documentclass{article}
\usepackage[T1]{fontenc}
\usepackage[utf8]{inputenc}
\usepackage{
ismir,
amsmath,
cite,
multirow,
tabularx,
graphicx,
color,
amsfonts,
enumitem,
booktabs,
tabularx,
footmisc,
xfrac,
enumitem,
float,
xurl,
soul,
}

\usepackage[bookmarks=false]{hyperref}
\hypersetup{colorlinks,allcolors=black}

\title{SurpriseNet: Melody Harmonization Conditioning on User-controlled Surprise Contours}

\multauthor 
{Yi-Wei Chen \hspace{1cm} Hung-Shin Lee \hspace{1cm} Yen-Hsing Chen \hspace{1cm} Hsin-Min Wang} {\bfseries \\ Institute of Information Science, Academia Sinica, Taiwan \\ 
{\tt\small chenvictor@iis.sinica.edu.tw}
}

\def\authorname{Yi-Wei Chen, Hung-Shin Lee, Yen-Hsing Chen, Hsin-Min Wang}

\begin{document}
\maketitle
\begin{abstract}

The surprisingness of a song is an essential and seemingly subjective factor in determining whether the listener likes it. With the help of information theory, it can be described as the transition probability of a music sequence modeled as a Markov chain. In this study, we introduce the concept of deriving entropy variations over time, so that the surprise contour of each chord sequence can be extracted. Based on this, we propose a user-controllable framework that uses a conditional variational autoencoder (CVAE) to harmonize the melody based on the given chord surprise indication. Through explicit conditions, the model can randomly generate various and harmonic chord progressions for a melody, and the Spearman's correlation and p-value significance show that the resulting chord progressions match the given surprise contour quite well. The vanilla CVAE model was evaluated in a basic melody harmonization task (no surprise control) in terms of six objective metrics. The results of experiments on the Hooktheory Lead Sheet Dataset show that our model achieves performance comparable to the state-of-the-art melody harmonization model.
\end{abstract}

\section{Introduction}

In recent years, deep learning has developed rapidly and has become the main technology for automatic music generation. In this study, we focus on the task of automatic melody harmonization, in which a system needs to assign appropriate and harmonic chords to a given melody. From previous studies, we have seen that the models based on the bidirectional long short-term memory (BiLSTM) perform well in this task \cite{Lim2017,Yeh2020}. Most of them can generate harmonic chords to harmonize a melody. In \cite{Sun2020}, by introducing blocked Gibbs sampling and class weighting, the model can further generate more reasonable and interesting chords, and is even comparable to human composers.

Based on these previous studies, we hope to further control the model to generate chords according to both melody conditions and user instructions. Latent representation learning is a powerful method that has been widely used in computer vision \cite{Kim2018,Chen2016,Li2018} and speech processing fields \cite{Hsu2017,Wang2018} to learn semantically meaningful information of attributes. Such techniques have also been applied to music processing in several recent works \cite{Tan2020,Brunner2018,Wang2020}. Motivated by these latent variable models, we modify the variational autoencoder (VAE) \cite{Sohn2015} so that the chord-to-chord mapping can be trained under the conditions of a melody sequence and a user-defined temporal contour. 

However, in the model, what are the most attractive and practical controllable conditions for users? Many high-level subjective emotions, such as happiness, sadness, surprisingness, and interestingness, can describe a musical sequence. These emotions seem to be abstract and difficult to quantify objectively. Fortunately, some of them can be calculated from the information dynamics \cite{Abdallah2009}. Previous music theory studies have shown that perceptual qualities and subjective states, such as uncertainty, surprisingness, complexity, tension, and interestingness, are closely related to the measurement of information theory, such as relative entropy and mutual information. In \cite{Abdallah2009}, the authors explored this idea in the context of Markov chains using musical sequences, resulting a structural analysis, which is largely consistent with the views of professional human listeners. Therefore, we use the surprisingness metric, which is defined as the negative log transition probability in a Markov chain, to generate the time-varying surprisingness contour of a chord sequence.

Similar to Mellotron, a text-to-speech system that uses pitch and rhythm contours as conditions to synthesize speech \cite{Valle2020}, in the training stage, we concatenate a chord sequence with additional conditions, namely its corresponding melody and surprise contour, feed them into an encoder to convert them into latent variables. Then, the latent variables are concatenated with the melody and surprise contour again, and input into a decoder to reconstruct the chord sequence. The model is expected to learn the latent representations of chords when the melody and surprise contour are given. Owing to the sampling mechanism of the VAE-based model, in the inference stage, we can randomly sample latent variables from the standard normal distribution to generate a variety of chords, which cohere the input melody and propagate according to the required surprise contour. In addition, we extend the 96 chords used in \cite{Sun2020} to all chord types in the Hooktheory Lead Sheet Dataset \cite{Anderson} to expand the chord selection of the model.

Our model is named SurpriseNet. It can harmonize a melody through user-controllable conditions. The highlights of SurpriseNet are two-fold. First, it relieves the tension between coherence and surprisingness caused by the melody and user-supplied condition, respectively. In general, it is easy to catch one factor (i.e., surprisingness) and lose another. Second, the results show that the vivid and harmonic chords generated by our model can not only correspond to the melody, but also strictly follow the given surprise contour. Several examples are available at \url{https://scmvp301135.github.io/SurpriseNet}.

\section{Related Work}

\subsection{Automatic Melody Harmonization}

Automatic melody harmonization aims to establish a learning model that can generate chord sequences to accompany a given melody \cite{Chuan2007,Simon2008}. In music, a chord is an arbitrary harmonic set consisting of three or more notes, which sounds as if these notes are sounding simultaneously \cite{Makris2016}. Conventionally, methods based on hidden Markov models (HMMs) \cite{Paiement2006,Tsushima2017,Tsushima2018} and genetic algorithms (GA) \cite{Kitahara2018} are commonly used to deal with the task.

Recently, some models based on deep learning have been proposed. For example, Lim \textit{et al.} first proposed a model based on the BiLSTM \cite{Lim2017}. The melody is input to the model to predict the simplified 24 chords(i.e., the major and minor triads) in the Wikifonia corpus. Based on the same model architecture in \cite{Lim2017}, Yeh \textit{et al.} proposed a model called MTHarmonizer, in which the chord types were extended to 48 by considering major, minor, augment and diminished chords in a larger corpus (i.e., the Hooktheory Lead Sheet Dataset) \cite{Yeh2020}. In addition, they integrated information of chord functions \cite{Chen2018} into the loss function to help chord label prediction \cite{Yeh2020}. However, there are several drawbacks in the above methods, such as overusing common chords and incorrect phrasing problems. Sun \textit{et al.} tried to solve these problems and produced interesting but still reasonable chords by introducing the orderless NADE training techniques, class weights, and Blocked Gibbs sampling into their model. They also extended the chord space to 96 types, including major, minor, augment, diminish, suspend, major7, minor7, and dominant7 \cite{Sun2020}.

\subsection{Controllable Music Generation}

Music generation can be regarded as a conditional estimation problem defined as $p(\text{music}|\text{condition})$, where both ``music'' and ``condition'' are usually time-series features. Related tasks include melody-based chord generation \cite{Simon2008} and chord-based melody generation \cite{Yang2017,Chen2018}. 

An alternative way is to learn the joint distribution $p(\text{music},\text{condition})$, and then set the condition during the generation process. Related tasks include automatic music completion or accompaniment based on the melody or chords \cite{Hadjeres2017,Zhu2018,Dong2018,Donahue2019,Simon2018}. However, many abstract music factors, such as music texture, melody contour, or other high-level subjective perception, are difficult to be explicitly encoded.

Latent representation learning is an ideal solution to the above problem, because representation learning embeds discrete music and condition sequences into a continuous latent space, and accurately captures the latent information from the music. Recent research has used disentangled representation learning to achieve controllable music generation models for style transfer, texture variation, and accompaniment arrangement \cite{Wang2020}. High-level subjective perception can also be captured in the latent space to generate music following the arousal condition \cite{Tan2020}. These studies show that VAEs \cite{Kingma2014,Roberts2018} are an effective framework for learning the representation of discrete music sequences. We incorporated this idea into our research, and expected the model to capture the latent information of chords when conditioned by melody sequences and surprise contours.

\section{Methodology}

In this section, we will introduce the calculation of surprise contours and the model architecture in detail. SurpriseNet is based on a conditional VAE, and its goal is to learn the representation of chords when conditioned by the melody sequences and surprise contours. In the inference process, the random latent variables, melody conditions, and surprise contours are provided to the decoder to produce harmonization. 

\subsection{Surprise Contour}

 Measures such as entropy and mutual information can be used to characterize random processes. One of the salient effects of listening to music is to create expectations for what is to come next, which may be fulfilled immediately, after a delay, or not at all depending on the situation. An essential aspect of this is that music is experienced as a phenomenon that ``unfolds'' over time, rather than being apprehended as a static object presented in its entirety \cite{Abdallah2009}. 
 
 Consider a snapshot of a stationary random process taken at a certain time: we can divide the timeline into the past and present parts, denoted as $t-1$ and $t$, respectively. Here we will consider one of the simplest random processes, a first-order Markov chain. Let S be a Markov chain with a finite state space $\{1, . . . ,N\}$ such that $S_t$ is the random variable representing the $t$-th element of the sequence. We can establish a transition matrix $a \in \mathbb{R}^{N\times N}$ encoding the distribution of any element of the chord sequence given the previous element, that is $p(\text{chord}_t|\text{chord}_{t-1}) = a_{t,t-1}$. According to the definition in \cite{Abdallah2009}, the surprise contour can be derived as the negative log probability as
\begin{equation}
\label{eqn:sur}
\text{Surprisingness} = -\log p(\text{chord}_t|\text{chord}_{t-1}).
\end{equation}

Equation (\ref{eqn:sur}) is actually the definition of the information content in information theory. It means that in a chord sequence, the higher the surprise value at a certain time, the greater the amount of information at that time.

\begin{figure}[t]
\begin{center}
\includegraphics[width=0.48\textwidth]{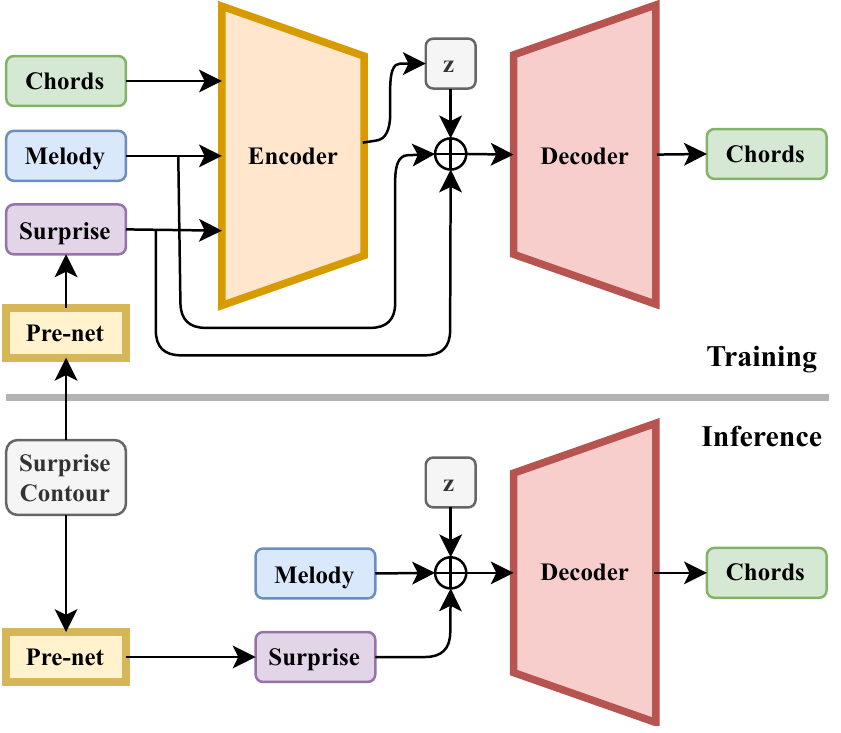}
\end{center}
\vspace{-20pt}
\caption{The structure of SurpriseNet. In the inference stage, only Decoder and Pre-net are used.}
\label{fig:surprisenet}
\vspace{-15pt}
\end{figure}

\subsection{VAE and Conditional VAE}

As described in \cite{Roberts2018}, a common goal for various kinds of autoencoders is that they compress the salient information within the input sample into a lower-dimensional latent code. Ideally, this will force the model to produce compact representations to capture the important factors of variation in the dataset. In pursuit of this goal, VAE \cite{Kingma2014,Rezende2014} introduces the constraint that the latent code $\mathbf{z}$ is a random variable distributed according to a prior $p(\mathbf{z})$.

The generative process of VAE is described as follows. A latent variable $\mathbf{z}$ is generated from the prior distribution $p(\mathbf{z})$, and the observation $\mathbf{x}$ is generated by the generative distribution $p(\mathbf{x}|\mathbf{z}$) conditioned on $\mathbf{z}$; that is, $\mathbf{z} \sim p(\mathbf{z})$ and $\mathbf{x} \sim p(\mathbf{x}|\mathbf{z})$. A VAE consists of an encoder $\theta$, which approximates the posterior $p(\mathbf{z}|\mathbf{x})$, and a decoder $\phi$, which parameterizes the likelihood $p(\mathbf{x}|\mathbf{z})$. Following the framework of variational inference, we do posterior inference by minimizing the KL divergence between the approximate posterior (i.e., the output of the encoder) and the true posterior $p(\mathbf{z}|\mathbf{x})$ by maximizing the evidence lower bound (ELBO). The objective function of VAE with Gaussian latent variables is
\begin{equation}
\tilde{\mathcal{L}}_{VAE} = \mathbb{E}[\log p_{\theta}(\mathbf{x}|\mathbf{z})] -KL(p_\phi(\mathbf{z}|\mathbf{x})||p(\mathbf{z})).
\end{equation}

In the common case where $p(\mathbf{z})$ is a diagonal-covariance Gaussian distribution, this can be circumvented by replacing $\mathbf{z} \sim \mathcal{N}(\bm{\mu}, \sigma\mathbf{I})$ with 
\begin{equation}
\mathbf{z} = \bm{\mu} + \bm{\sigma} \odot \epsilon,
\end{equation}
where $\epsilon$ is a small random factor.

As for the conditional VAE \cite{Sohn2015}, the conditional generative process of the model is given as follows. For a given condition $\mathbf{c}$, $\mathbf{z}$ is drawn from the prior distribution $p(\mathbf{z}|\mathbf{c})$ realized by a standard Gaussian distribution, and the output $\mathbf{x}$ is generated from the distribution $p_\theta(\mathbf{x}|\mathbf{z},\mathbf{c})$. Therefore, the training objective function can be expressed as
\begin{equation}
\tilde{\mathcal{L}}_{CVAE} = \mathbb{E}[\log p_{\theta}(\mathbf{x}|\mathbf{z},\mathbf{c})] -KL(p_\phi(\mathbf{z}|\mathbf{x},\mathbf{c})||p(\mathbf{z}|\mathbf{c})).
\end{equation}

Our work is divided into two parts. We first train the conditional VAE models for 96 or 633 types of chords, conditioned only on the melody, to observe the performance in completing the basic harmonization task. Next, we select the better conditional VAE to train the final SurpriseNet in combination with the surprise contours.

\begin{figure}[t]
\begin{center}
\includegraphics[width=0.48\textwidth]{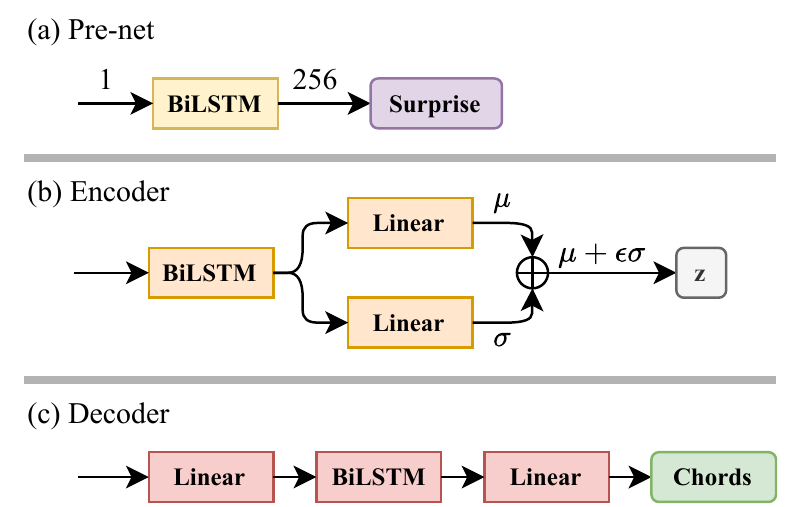}
\end{center}
\vspace{-15pt}
\caption{Three main components of SurpriseNet.}
\label{fig:components}
\vspace{-15pt}
\end{figure}

The main architecture of SurpriseNet and its key components are shown in Fig. \ref{fig:surprisenet} and Fig. \ref{fig:components}, respectively. Our components of VAE follow the structures used in \cite{Bowman2016} and MusicVAE \cite{Roberts2018}, and finally combine with the conditional part \cite{Sohn2015} to complete the harmonization task. In the training stage, the surprise contour is processed by the Pre-net implemented by a BiLSTM, as shown in Fig. \ref{fig:components}(a), to extend the feature from a scalar to a 256-dimensional vector. Afterwards, the features of the chord, melody, and extended surprise contour are concatenated and fed into the encoder to generate a latent code $\mathbf{z}$ subject to a standard normal distribution. The encoder is also implemented by a BiLSTM, as shown in Fig. \ref{fig:components}(b), where the size of each layer is 256 or 512. Different from the RNN-based VAE \cite{Bowman2016} and MusicVAE \cite{Roberts2018}, the frame-wise outputs are further transformed by two linear layers to respectively generate the sequential latent variables, $\bm\mu$ and $\bm\sigma$, with dimensionalities of 16 or 64.

As for the decoder, instead of performing it in an autoregressive manner as in MusicVAE, we concatenate $\mathbf{z}$, melody, and extended surprise representation frame by frame as inputs of the decoder to reconstruct the chord sequence. The number of layers and hidden size in the decoder are the same as those in the encoder. Dropout was employed with a rate of 0.2 on each BiLSTM to prevent overfitting. The batch size was set to 64. Early stopping for 10 epoch patience was applied.

In the inference stage, given the melody and the surprise contour, the surprise contour is first processed by the Pre-net. Then, the latent variable is sampled from a standard normal distribution. Finally, the latent variable, melody, and extended surprise representation are input to the decoder to complete the harmonization process. The implementation details of the model are available at \url{https://github.com/scmvp301135/SurpriseNet}.

\section{Experiments}

\subsection{Datasets}

We performed experiments on the Hooktheory Lead Sheet Dataset (HLSD) \cite{Anderson}, which contains high-quality and human-arranged melodies with chord progressions. The dataset is provided in two formats, event-based JSON files and MIDI files. Furthermore, there are many types of labels on chords, such as chord symbols and Roman numerals for reference. The dataset contains a total of 633 chord types, including 9th, 11th, 13th, half diminished, and slash chords. In previous studies conducted on the dataset \cite{Lim2017,Yeh2020,Sun2020}, the number of chord types was simplified to 48 (major and minor triads) or 96 (major, minor, augment, diminish, suspend, major7, minor7,
and dominant7). In this study, we experimented with two settings, 96 and 633 chord types.

We followed the data split in \cite{Sun2020}; the training set contains 17,505 samples, the test set contains 500 samples. For each song, the melody and chords are aligned every two beats in a measure, and the chords are encoded into a one-hot format for training.

\subsection{Objective Metrics}

For objective evaluation, we used six different objective metrics proposed in \cite{Yeh2020}. The first three metrics measure the quality of the chord progression, and the others measure the harmonicity between the melody and the chords.

\begin{itemize}[leftmargin=*]
\setlength\itemsep{-0.3em}
\item {\bf{Chord histogram entropy (CHE):}} 
The entropy of the chord histogram.

\item {\bf{Chord coverage (CC):}}
The number of chord types in a chord sequence.

\item {\bf{Chord tonal distance (CTD):}}
The tonal distance between two chords when they are represented by 6-D feature vectors \cite{Harte2006}.

\item {\bf{Chord tone to non-chord tone ratio (CTnCTR):}}
The ratio of the number of chord tones to the number of non-chord tones.

\item {\bf{Pitch consonance score (PCS):}}
The sum of the consonance scores between a melody note and each note in a given chord. 

\item {\bf{Melody-chord tonal distance (MCTD):}}
The tonal distance between a melody note and a chord when they are represented by 6-D feature vectors.

\end{itemize}

\subsection{Surprise Contour Evaluation}

The task of user-controlled melody harmonization has never been seen in the literature. Therefore, there is no baseline systems for comparison. We decided to use some statistical methods to evaluate the correlation or causation between a given contour and the generated sample, instead of subjective testing.

Unlike the pitch and rhythm error evaluation in Mellotron \cite{Valle2020}, we used Spearman's correlation, which is suitable for continuous and discrete ordinal values between two variables. The p-value was used to determine the significance of the results under the assumption that there is no significant correlation between the surprisingness trend of the predicted chord progression and the given surprise contour. A small p-value indicates strong evidence against the assumption, which means that the results are correlated to some extent. 

The reason for not using error evaluation, such as the mean squared error (MSE), is that the harmonization result is jointly decided by the melody and the surprise contour. If the error between the generated trend and the given surprise contour is zero, it means that the model completely follows the surprise contour and ignores the melody conditions. Obviously, this is not acceptable and will lead to discordant harmonization results.

\section{Results}

In this section, we will first compare the conditional VAE models with Sun \textit{et al.}'s model \cite{Sun2020} and human performance in terms of six objective metrics. Then, we will illustrate some harmonization samples generated by SurpriseNet based on different surprise contours. The last part is the correlation analysis. 

\subsection{Objective Evaluation}

\begin{table}[t!]
\vspace{-8pt}
\caption {Objective evaluation results with respect to various models. For the metrics related to Chord Progression, the higher value in CHE and CC means the higher diversity of the generated chords, and the lower value in CTD implies that the chord progression is smoother. As for the metrics related to Harmonicity, the higher value in CTnCTR and PCS and the lower value in MCTD indicate better harmonization results. The arrow denotes whether the metric is the larger the better or the lower the better.}
\vspace{10pt}
\label{table:objective}
\centering
\begin{tabular*}{\linewidth} {@{\extracolsep{\fill}} cccc}
\toprule
\multicolumn{1}{l} {\textbf{Chord Progression}}  & \hspace{-12pt} CHE$\uparrow$ & \hspace{-12pt} CC$\uparrow$ & \hspace{-12pt} CTD$\downarrow$ \\ 
\midrule
\multicolumn{1}{l} {Humans} & \hspace{-12pt} $1.266$ & \hspace{-12pt} $4.344$ & \hspace{-12pt} $0.628$ \\
\multicolumn{1}{l} {Sun \textit{et al.}, $|S|=96$} & \hspace{-12pt} $1.280$ & \hspace{-12pt} $4.900$ & \hspace{-12pt} $0.730$ \\
\multicolumn{1}{l} {CVAE, $|S|=96$} & \hspace{-12pt} $1.210$ & \hspace{-12pt} $4.712$ & \hspace{-12pt} $\bf{0.577}$ \\
\multicolumn{1}{l} {CVAE weight, $|S|=96$} & \hspace{-12pt} $1.670$ & \hspace{-12pt} $7.360$ & \hspace{-12pt} $0.620$ \\
\multicolumn{1}{l} {CVAE, $|S|=633$} & \hspace{-12pt} $1.644$ & \hspace{-12pt} $7.074$ & \hspace{-12pt} $0.730$ \\
\multicolumn{1}{l} {CVAE weight, $|S|=633$} & \hspace{-20pt} $\bf{1.934}$ \hspace{-12pt} & \hspace{-12pt} $\bf{9.890}$ & \hspace{-12pt} $0.649$ \\
\midrule
\midrule
\multicolumn{1}{l} {\textbf{M/C Harmonicity}} & \hspace{-12pt} CTnCTR$\uparrow$ & \hspace{-12pt} PCS$\uparrow$ & \hspace{-12pt} MCTD$\downarrow$   \\
\midrule
\multicolumn{1}{l} {Humans} & \hspace{-12pt} $0.726$ & \hspace{-12pt} $0.515$ & \hspace{-12pt} $1.276$ \\
\multicolumn{1}{l} {Sun \textit{et al.}, $|S|=96$} & \hspace{-12pt} $\bf{0.887}$ & \hspace{-12pt} $\bf{0.652}$ & \hspace{-12pt} $\bf{1.052}$ \\
\multicolumn{1}{l} {CVAE, $|S|=96$} & \hspace{-12pt} $0.851$ & \hspace{-12pt} $0.611$ & \hspace{-12pt} $1.110$ \\
\multicolumn{1}{l} {CVAE weight, $|S|=96$} & \hspace{-12pt} $0.841$ & \hspace{-12pt} $0.523$ & \hspace{-12pt} $1.190$ \\
\multicolumn{1}{l} {CVAE, $|S|=633$} & \hspace{-12pt} $0.767$ & \hspace{-12pt} $0.530$ & \hspace{-12pt} $1.229$ \\
\multicolumn{1}{l} {CVAE weight, $|S|=633$} & \hspace{-12pt} $0.705$ & \hspace{-12pt} $0.476$ & \hspace{-12pt} $1.290$ \\
\bottomrule
\end{tabular*}
\vspace{-10pt}
\end{table}

The objective evaluation results are shown in Table \ref{table:objective}. Compared with the results of humans and Sun \textit{et al.}'s model, we can see that the vanilla CVAE model without using class weights (cf. CVAE, $|S|=96$) achieved comparable results in all metrics. It performed better in CTD, indicating that the generated chord progression is smoother. After introducing chord balancing (cf. CVAE weight, $|S|=96$), the CVAE model learned how to sample rare chords, thereby increasing the chord diversity, as shown in the results in CC and CHE.

We further expanded the chord space to 633 to maintain the integrity of the chords in the dataset. Due to the extension of the chord dimension, the deeper CVAE without using class weights (cf. CVAE, $|S|=633$) obtained results comparable to the vanilla CVAE model with chord balancing (cf. CVAE weight, $|S|=96$). After introducing chord balancing (CVAE weight, $|S|=633$), the deeper CVAE model achieved the best chord diversity, with an average of nearly 10 chord types in a musical sequence.

Despite the above improvements, trade-offs in other metrics (such as CTnCTR, PCS, and MCTD) can be observed. As pointed out in \cite{Sun2020}, rare chords, such as 7th, 9th, 11th, and 13th, will cause a degradation in the melody/chord harmonicity metrics, but this is mainly due to the definition of CTnCTR, PCS, and MCTD. In order to maintain the diversity of chords, we decided to train SurpriseNet with the conditional VAE architecture considering 633 chord types.

\begin{figure}[t]
\begin{center}
\includegraphics[width=0.48\textwidth]{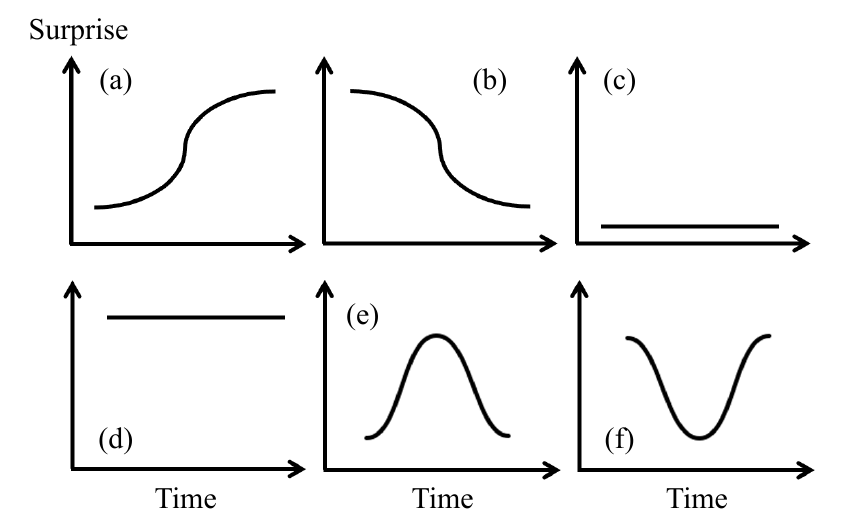}
\end{center}
\vspace{-20pt}
\caption{The six given surprise contours.}
\label{fig:uc}
\vspace{-10pt}
\end{figure}

\subsection{Generated Samples}

We compared the chord generation results of SurpriseNet (based on CVAE, $|S|=633$) and weighted SurpriseNet (CVAE weight, $|S|=633$) based on 6 representative surprise contours, as shown in Fig. \ref{fig:uc}. The 6 contours were generated by the sigmoid function, plain line, and normal distribution, and their reversed profiles, respectively. We intended to check whether the generated chord progression really follows the surprise contour to harmonize the given melody.

To use the sigmoid function to represent the surprise contour, we first normalized it to match the maximum value in the surprise contours of the training data. This contour (cf. Fig. \ref{fig:uc}(1)) represents a song with lower chord variation in the first half and higher chord variation in the second half. The reverse sigmoid function leads to the opposite trend (cf. Fig. \ref{fig:uc}(2)). In the case of plain line, we used two sequences consisting of zero (cf. Fig. \ref{fig:uc}(3)) and the maximum value (cf. Fig. \ref{fig:uc}(4)) as the surprise contours. A surprise contour with all values being zero indicates that there should be no fluctuation in the chord sequence. In other words, it is expected to see that the given melody will always be harmonized with the same chord. As for the surprise contour with all values being the maximum, it indicates that there should be a lot of up-and-down changes in the resulting chord sequence. That is, it is expected to see that the given melody will be harmonized with various chords. These two cases are considered the most extreme cases. According to the normal distribution, we expect that the highest arousal will appear in the middle of the harmonization result (cf. Fig. \ref{fig:uc}(5)). As for its inverse profile (cf. Fig. \ref{fig:uc}(6)), it is expected to generate a plain and more predictable result in the middle of the chord sequence.

\begin{figure}[t]
\vspace{-30pt}
\begin{center}
\includegraphics[width=0.5\textwidth]{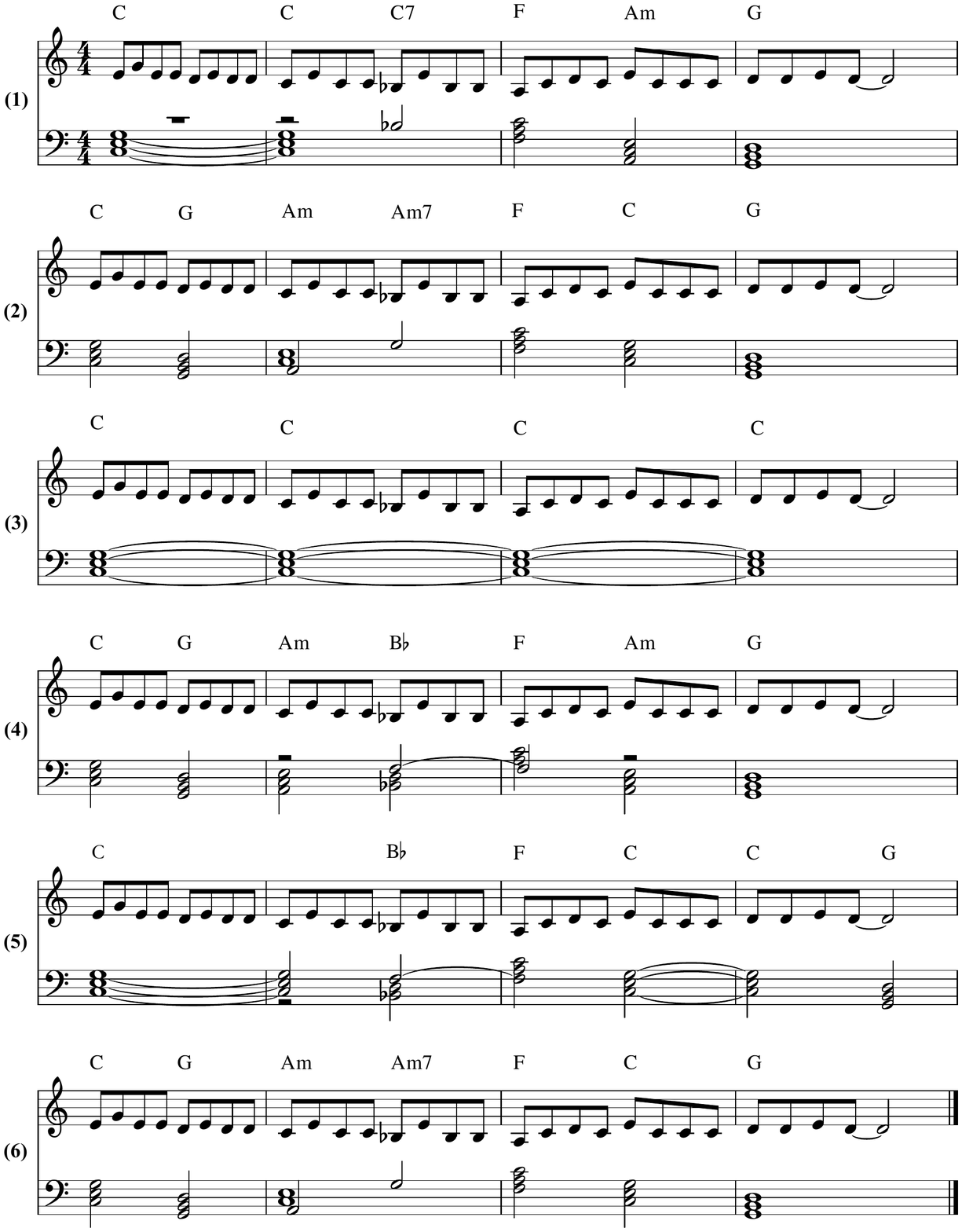}
\end{center}
\vspace{-40pt}
\caption{Samples generated by SurpriseNet.}
\label{fig:sn}
\end{figure}

\begin{figure}[t]
\vspace{-30pt}
\begin{center}
\includegraphics[width=0.5\textwidth]{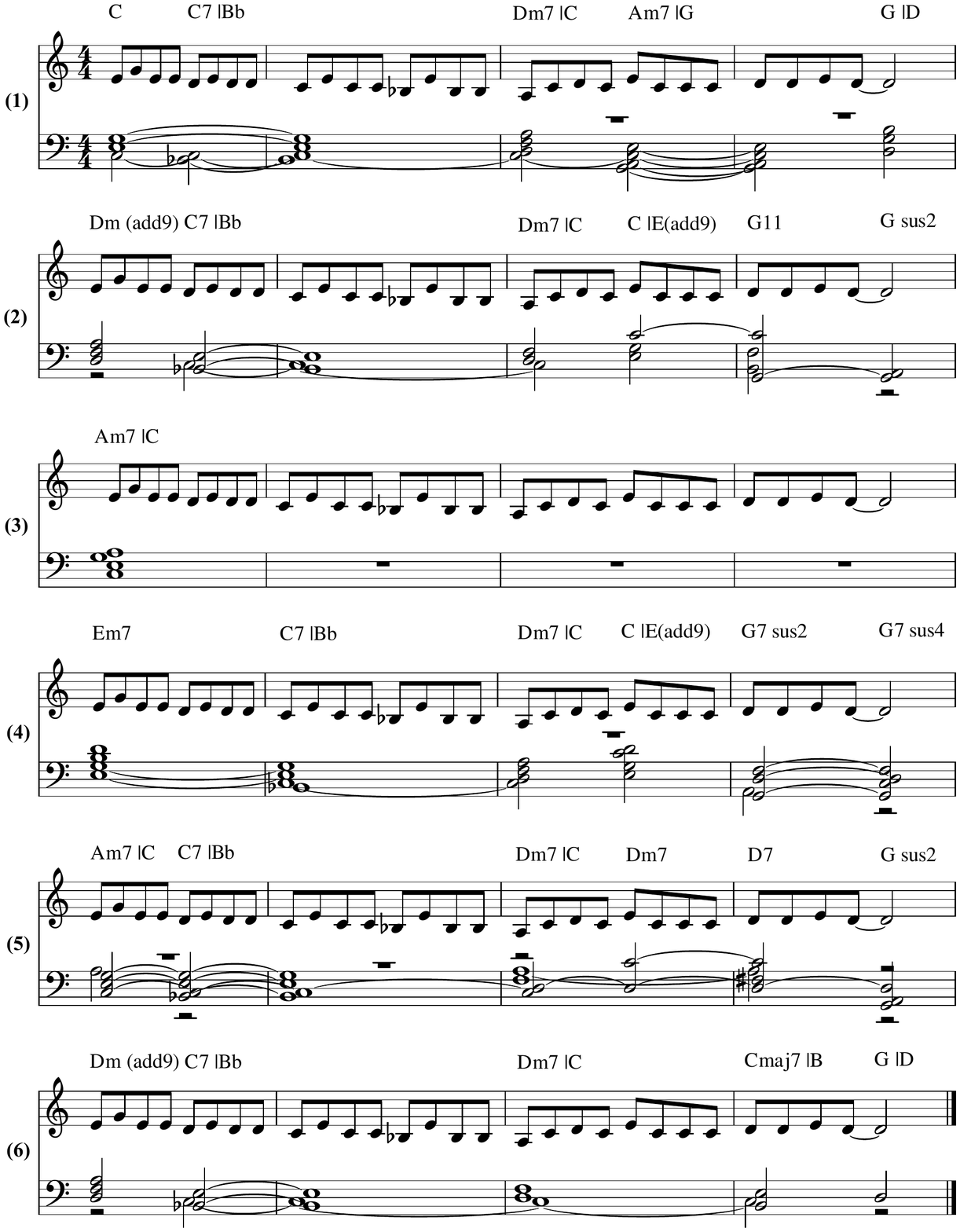}
\end{center}
\vspace{-40pt}
\caption{Samples generated by weighted SurpriseNet.}
\label{fig:snc}
\end{figure}

Fig. \ref{fig:sn} and Fig. \ref{fig:snc} show the harmonization results of SurpriseNet and weighted SurpriseNet for a 4-measure melody, respectively. They are displayed in the same order as the function types in Fig. \ref{fig:uc}.
From Fig. \ref{fig:sn}(1), as expected, we can see that SurpriseNet generated continuous C chords for the first two measures at the beginning, and then generated varying chords for the last part of the song, according to the given sigmoid-like surprise contour in Fig. \ref{fig:uc}(1). From Fig. \ref{fig:sn}(2), we can also see that SurpriseNet generated different chords at the beginning, and then generated more C and G chords that appeared previously, following the given reverse sigmoid-like surprise contour in Fig. \ref{fig:uc}(2). As for the results of weighted SurpriseNet (see Figs. \ref{fig:snc}(1) and \ref{fig:snc}(2)), the chord progressions generated were not exactly as expected, but some surprising and complicated chords were brought in for users' reference. Next, given the all-zero surprise contour in Fig. \ref{fig:uc}(3), it is obvious that both models followed the condition to generate only one type of chord in the results (see Figs. \ref{fig:sn}(3) and \ref{fig:snc}(3)). As for the all-maximum surprise contour in Fig. \ref{fig:uc}(4), as shown in Figs. \ref{fig:sn}(4) and \ref{fig:snc}(4), it is also obvious that the results generated by the two models changed in the chord type almost every two beats, which is the minimum time unit for changing the chord in the training data. For the normal distribution contour in Fig. \ref{fig:uc}(5), the result generated by SurpriseNet is roughly as expected, with more chord changes in the middle of the song (see Figs. \ref{fig:sn}(5)). But the result of weighted SurpriseNet is quite different from expectations (see Figs. \ref{fig:snc}(5)). For the inverse normal distribution contour in Fig. \ref{fig:uc}(6), the results of both models are not in line with our expectations. But we can see that they are similar to the results generate based on the reverse sigmoid-like surprise contour in the beginning part of the song. When the model is initially assigned a high surprise value, the trend in the first part of the output is similar.

In summary, the above samples generated by SurpriseNet are almost in good agreement with our expectations. The model can indeed generate chords that have a tendency to follow a given surprise contour. However, weighted SurpriseNet seems to over concentrate on using more complicated or rare chords to harmonize the melody due to the class penalty (i.e., weights), so that the trend is not clearly consistent with the given contour. 

\subsection{Correlation Measurement}

\begin{table}[t]
\vspace{-10pt}
\caption {Spearman's $\rho$ and p-value significance of SurpriseNet and weighted SurpriseNet.}
\vspace{10pt}
\label{table:corr}
\centering
\begin{tabular*}{0.9\linewidth} {@{\extracolsep{\fill}} ccc}
\toprule
\multicolumn{1}{c} {\bf Method} & \bf Spearman's $\rho$  & \bf p-value \\
\midrule
\multicolumn{1}{c} {SurpriseNet} & $0.517$ & $< 0.001$  \\
\multicolumn{1}{c} {Weighted SurpriseNet} & $0.406$ & $<0.001$ \\
\bottomrule
\end{tabular*}
\vspace{-15pt}
\end{table}

Because there is no existing model for this task, we use Spearman's $\rho$ and p-value to evaluate the correlation and significance between the surprisingness values in the given surprise contour and the surprisingness values in the generated chord progression. 
From Table \ref{table:corr}, we can find that there is indeed a certain correlation between the given surprise contour and the generated chord progression. Furthermore, SurpriseNet seems to be more controllable than weighted SurpriseNet, with a higher Spearman's $\rho$. The p-value shows that there is no significant difference between the surprisingness trend of the given surprise contour and the surprisingness trend of the generated chord progression for the two models. The result implicitly confirms that given a surprise contour and a melody, these models can generate the corresponding chords as instructed to complete the melody harmonization task, thereby achieving a user-controlled model.

\section{Future Work}

The HLSD dataset contains rich intonation data, such as Roman, symbol, secondary, and mode data. We can introduce some approaches from the NLP field, such as modeling these data by referring to various language models. Moreover, in this work, the rhythmic type of chords is simplified and restricted to two beats in a measure. But in fact, there are various rhythmic types in the dataset, such as syncopation and tuplets. Perhaps a disentangled representation learning model can be used to capture this information, so that we can implement an omni model with more complicated rhythms.

In addition, surprise is still an open for discussion topic. In this work, we only consider the surprise in the chord sequence. In the future, can also consider the surprise in the melody sequence at the same time. Moreover, the surprise can be considered not only as the conditional probability of past chord events but also as the conditional probability of the melody sequence at that time. These different considerations will bring different meanings to the surprise.

\section{Conclusions}

In this paper, we proposed SurpriseNet, which is based on a conditional VAE model and combines a surprise contour from the transition probability in a Markov chain, to achieve a user-controlled melody harmonization task. From the generated samples, we observed that the model could accurately generate various harmonic chord progressions according to the given surprise contours. The Spearman's correlation and p-value significance show that there is a positive correlation between the given surprise contour and the generated chord progression.

The vanilla conditional VAE model was evaluated in the basic melody harmonization task. The objective evaluation results show that the conditional VAE model could achieve performance comparable to the state-of-the-art melody harmonization model \cite{Sun2020}. We expanded the chord space from 96 to 633 to broaden the range of chord selection for the model. The conditional VAE model could generate more types of chords, such as 7th, 9th, 11, 13th, and slash chords, resulting in vivid and harmonic results.

\bibliography{references.bib}

\end{document}